\documentclass[12pt]{article}

\usepackage{graphics}
\usepackage{epsfig}
\usepackage{cite}
\input epsf

\title{Phenomenology of $k_T$-factorization for inclusive top quark pair production at hadron colliders}

\author{A.V.~Lipatov, N.P.~Zotov}

\begin{document}

\maketitle

\begin{center}

{\it D.V.~Skobeltsyn Institute of Nuclear Physics,\\ 
M.V. Lomonosov Moscow State University,
\\119991 Moscow, Russia\/}

\end{center}

\vspace{1cm}

\begin{center}

{\bf Abstract }

\end{center}

We investigate the inclusive top quark pair production in proton-proton and proton-antiproton collisions
at high energies in the framework of $k_T$-factorization QCD approach. Our study is
based on the off-shell partonic subprocesses $g^* g^* \to Q \bar Q$ and $q^* {\bar q}^* \to Q \bar Q$,
where the transverse momentum of both incoming quarks and gluons are taken into account.
The unintegrated parton densities in a proton are determined by the Kimber-Martin-Ryskin prescription
as well as the CCFM evolution equation. The conservative error analisys is performed and 
comparison with the results of traditional NLO pQCD calculations is done.
Our predictions agree well with the recent experimental data taken by the D0, CDF, CMS and ATLAS 
collaborations at the Tevatron and LHC energies.

\vspace{1.5cm}

\noindent
PACS number(s): 14.65.Ha, 12.38.-t

\newpage

With the startup of the LHC, high energy particle physics entered a new era. 
One of the important part of the LHC program are the investigations
of top-antitop pair production\cite{1,2}. These studies are interesting for several reasons.
At higher (compared to the Tevatron) energies, measurements with higher precision become available, 
which provide an opportunity to test the predictions of various theoretical models. 
The top quark mass $m_t$ is an important parameter of the Standard Model (SM) and it affects predictions 
of other SM observables via radiative corrections. A precise measurement of $m_t$
is crucial since it is one of the most important inputs to the global electroweak fits
which provide constraints on the SM itself, including indirect limits on the mass of the
undiscovered Higgs boson. 
Moreover, top pair production is an important background in various
searches for physics beyond the SM. 
Due to the large mass of the top quark,
many models of physics beyond the SM predict observable effects in the top quark sector 
which can affect the top quark production rate. Finally, the abundant $t\bar t$ sample can be 
used for improving many aspects of detector performance.

From the theoretical point of view, $t \bar t$ quark pairs 
in $pp$ and $p\bar p$ collisions are produced via 
standard QCD gluon-gluon fusion and quark-antiquark annihilation
subprocesses. At the LHC, the production mechanism is dominated
by the gluon-gluon fusion, whereas at the Tevatron, top quark pairs are predominantly produced 
through quark-antiquark annihilation. 
In the framework of standard QCD, $t\bar t$ pair production at modern energies
has been studied in many papers (see, for example,\cite{3,4,5,6,7,8,9} and references therein).
In particular, theoretical calculations of $t\bar t$ cross section 
have been carried out at next-to-leading order (NLO)\cite{7,8} and approximate 
next-to-next-to-leading order (NNLO)\cite{9} of QCD. The results of these calculations
agree with the Tevatron and LHC data within the theoretical and experimental uncertainties.
However, an alternative description can be provided by the $k_T$-factorization approach of QCD\cite{10}. 
%In this approach, the large logarithmic contributions to the cross section 
%proportional to $\ln s \sim \ln 1/x$ are summed up to all orders of perturbation theory 
%(in the leading logarithmic approximation) using famous Balitsky-Fadin-Kuraev-Lipatov (BFKL)\cite{6} 
%or Ciafaloni-Catani-Fiorani-Marchesini (CCFM)\cite{7} equations.
A detailed description and discussion of the $k_T$-factorization formalism can be
found, for example, in reviews\cite{11}. Here we only mention that this
approach has been successfully applied recently to describe the heavy flavour\cite{12, 13} and prompt
photon\cite{14} production at the Tevatron and LHC energies.

The present note is motivated by very recent measurements\cite{1,2} performed by the 
CMS and ATLAS collaborations where the $t \bar t$ pair cross section in $pp$ collisions at $\sqrt s = 7$~TeV
has been measured. Our main goal is to give a systematic analysis of first
LHC data\cite{1,2} as well as recent Tevatron data\cite{15,16,17} in the framework of $k_T$-factorization 
approach. We consider both gluon-gluon fusion and quark-antiquark annihilation 
subprocesses and take into account the non-zero transverse momentum of incoming gluons and quarks 
in a proper way. Specially we study different sources of theoretical uncertainties. 
Such calculations are performed for the first time.

The cross section of $t\bar t$ pair production in the $k_T$-factorization approach
is calculated as a convolution of off-shell ($k_T$-dependent)
partonic cross sections and corresponding 
unintegrated parton distributions in a proton. The contributions 
from the gluon-gluon fusion and quark-antiquark
annihilation can be presented in the following simple forms:
$$
  \displaystyle \sigma_{gg} = \int {1\over 16\pi (x_1 x_2 s)^2 } f_g(x_1,{\mathbf k}_{1T}^2,\mu^2) f_g(x_2,{\mathbf k}_{2T}^2,\mu^2) |\bar {\cal M}(g^* g^* \to t\bar t)|^2 \times \atop
  \displaystyle  \times d{\mathbf p}_{1T}^2 d{\mathbf k}_{1T}^2 d{\mathbf k}_{2T}^2 dy_1 dy_2 {d\phi_1 \over 2\pi} {d\phi_2 \over 2\pi}, \eqno (1)
$$
$$
  \displaystyle \sigma_{q\bar q} = \sum_q \int {1\over 16\pi (x_1 x_2 s)^2 } f_q(x_1,{\mathbf k}_{1T}^2,\mu^2) f_q(x_2,{\mathbf k}_{2T}^2,\mu^2) |\bar {\cal M}(q^* {\bar q}^* \to t\bar t)|^2 \times \atop
  \displaystyle  \times d{\mathbf p}_{1T}^2 d{\mathbf k}_{1T}^2 d{\mathbf k}_{2T}^2 dy_1 dy_2 {d\phi_1 \over 2\pi} {d\phi_2 \over 2\pi}, \eqno (2)
$$

\noindent 
where $f_q(x,{\mathbf k}_{T}^2,\mu^2)$ and $f_g(x,{\mathbf k}_{T}^2,\mu^2)$ are the unintegrated quark and 
gluon distributions in a proton, 
$|\bar {\cal M}(g^* g^* \to t\bar t)|^2$ and $|\bar {\cal M}(q^* {\bar q}^* \to t\bar t)|^2$ are the 
off-shell (depending on the ${\mathbf k}_{1T}^2$ and 
${\mathbf k}_{2T}^2$) matrix elements squared, and 
$s$ is the total center-of-mass energy.
The produced top quark and anti-quark have the 
transverse momenta ${\mathbf p}_{1T}$ and
${\mathbf p}_{2T}$ and the center-of-mass rapidities $y_1$ and $y_2$.
The initial off-shell partons have fractions $x_1$ and $x_2$ 
of parent protons longitudinal 
momenta, non-zero transverse momenta ${\mathbf k}_{1T}$ and 
${\mathbf k}_{2T}$ and azimuthal angles $\phi_1$ and $\phi_2$. 
The evaluation of $|\bar {\cal M}(g^* g^* \to Q\bar Q)|^2$, where $Q$ is any heavy quark, 
is described in detail in\cite{18}.
The evaluation of $|\bar {\cal M}(q^* {\bar q}^* \to Q\bar Q)|^2$ is straightforward.
Here we only mention two technical points.
First, according to the $k_T$-factorization prescription\cite{10},
the summation over the incoming off-shell gluon polarizations in (1) is 
carried with $\sum \epsilon^\mu \epsilon^{  \, \nu} = {\mathbf k}_T^{\mu} {\mathbf k}_T^{\nu}/{\mathbf k}_T^2$,
where ${\mathbf k}_T$ is the gluon transverse momentum.
Second, when we calculate the squared matrix element in (2),
the spin density matrix for off-shell spinors is taken in the
form $u (p) \bar u (p) = x \hat p_p$\cite{19}, where $x$ is the fraction of 
initial proton longitudinal momentum $p_p$. In all other respects our calculations follow the 
standard Feynman rules. Note that since the expression for the off-shell quark spin density 
matrix has been derived in the massless approximation, numerically we neglect the masses of any incoming quarks. 

To determine unintegrated quark and gluon densities in 
a proton we apply the Kimber-Martin-Ryskin (KMR) approximation\cite{20}. This approach is the prescription
to construct the unintegrated parton distributions from the known conventional ones. 
In this approximation, the unintegrated quark and 
gluon distributions are given by
$$
  \displaystyle f_q(x,{\mathbf k}_T^2,\mu^2) = T_q({\mathbf k}_T^2,\mu^2) {\alpha_s({\mathbf k}_T^2)\over 2\pi} \times \atop {
  \displaystyle \times \int\limits_x^1 dz \left[P_{qq}(z) {x\over z} q\left({x\over z},{\mathbf k}_T^2\right) \Theta\left(\Delta - z\right) + P_{qg}(z) {x\over z} g\left({x\over z},{\mathbf k}_T^2\right) \right],} \eqno (3)
$$
$$
  \displaystyle f_g(x,{\mathbf k}_T^2,\mu^2) = T_g({\mathbf k}_T^2,\mu^2) {\alpha_s({\mathbf k}_T^2)\over 2\pi} \times \atop {
  \displaystyle \times \int\limits_x^1 dz \left[\sum_q P_{gq}(z) {x\over z} q\left({x\over z},{\mathbf k}_T^2\right) + P_{gg}(z) {x\over z} g\left({x\over z},{\mathbf k}_T^2\right)\Theta\left(\Delta - z\right) \right],} \eqno (4)
$$
\noindent
where $P_{ab}(z)$ are the usual unregulated LO DGLAP splitting 
functions. The theta functions which appear 
in~(3) and~(4) imply the angular-ordering constraint $\Delta = \mu/(\mu + |{\mathbf k}_T|)$ 
specifically to the last evolution step to regulate the soft gluon
singularities. The Sudakov form factors $T_q({\mathbf k}_T^2,\mu^2)$ and 
$T_g({\mathbf k}_T^2,\mu^2)$ enable us to include logarithmic loop corrections
to the calculated cross sections. 
Numerically, for the input we have used leading-order parton densities $q(x,\mu^2)$ and 
$g(x,\mu^2)$ from recent MSTW set\cite{21}. 

Another solution for the unintegrated gluon distribution
have been obtained\cite{22} from the CCFM evolution 
equation where all input parameters 
have been fitted to describe the proton structure function $F_2(x, Q^2)$.
Below we will use proposed gluon density (namely, set A0)
to study the uncertainties of our calculations
connected with the non-collinear evolution. Since we take into account the quark-antiquark
annihilation subprocess, we should apply relevant unintegrated quark distributions to accomplish the 
CCFM gluon ones. At present, however, unintegrated quark densities are available in
the framework of KMR approach only\footnote{See, for example, reviews\cite{11} for more details.}, 
therefore we will use following simple approximation. We will divide the quark densities into 
several parts which correspond to the interactions of valence quarks, sea quarks appearing at the last 
splitting and sea quarks coming from the earlier (second-to-last, third-to-last and other)
steps of gluon evolution.
First of them, unintegrated valence quark distributions have been obtained\cite{23} from the 
numerical solution of the CCFM-like 
equation\footnote{Authors are very grateful to Hannes Jung for providing us the code
for the unintegrated valence quark distributions.}. 
To calculate the contribution of the
sea quarks appearing at the last step of 
the gluon evolution we convolute the CCFM-evolved unintegrated gluon distribution 
$f_g(x,{\mathbf k}_T^2,\mu^2)$ with the standard leading-order DGLAP splitting function 
$P_{qg}(z)$:
$$
  f_q^{(sea)}(x,{\mathbf k}_T^2,\mu^2) = {\alpha_s({\mathbf k}_T^2)\over 2 \pi} \int\limits_x^1 f_g(x/z,{\mathbf k}_T^2,\mu^2)P_{qg}(z)\,dz. \eqno(5)
$$

\noindent
Note that in the region of small ${\mathbf k}_T^2 < q_0^2$ the
scale in the strong coupling constant $\alpha_s$ is kept to be fixed at $q_0 = 1$~GeV.
To estimate the contribution of the sea quarks 
coming from the earlier evolution steps we apply the procedure based on the specific properties of the KMR 
scheme. Modifying~(3) in such a way\cite{24} that only the first term is 
kept and the second term is omitted and keeping only the sea quark in first term, we remove the 
valence and $f_q^{(sea)}(x,{\mathbf k}_T^2,\mu^2)$ quarks from the evolution ladder.
In this way contribution of the sea quarks coming from the earlier gluon splittings to
the $f_q(x,{\mathbf k}_T^2,\mu^2)$ is taken into account only. 

Other essential parameters were taken as follows:
top quark mass $m_t = 170$~GeV, renormalization and factorization scales $\mu = \xi m_t$ (where
we vary the parameter $\xi$ between 1/2 and 2 about the default value $\xi = 1$
in order to estimate the scale uncertainties of our calculations),
LO formula for the strong coupling constant $\alpha_s(\mu^2)$ 
with $n_f = 4$ massless quark flavours and $\Lambda_{\rm QCD} = 200$ MeV, 
such that $\alpha_s(M_Z^2) = 0.1232$.
The multidimensional integration in~(1) and~(2) has been performed
by means of the Monte Carlo technique, using the routine \textsc{vegas}\cite{25}.
The full C$++$ code is available from the authors on 
request\footnote{lipatov@theory.sinp.msu.ru}.

We now are in a position to present our numerical results.
In Fig.~1 we confront the calculated top quark pair total and differential 
(as a function of $t$-quark transverse 
momentum) cross sections in $p\bar p$ and $pp$ collisions 
with the recent data\cite{1,2,15,16,17,26,27} taken by the D0, CDF, CMS and ATLAS 
collaboration at the Tevatron and LHC. For comparison we plot 
also the NLO pQCD predictions\cite{8} listed in\cite{15}.
One can see that the experimental data are reasonable well described by
the $k_T$-factorization approach. 
Concerning the transverse momentum distribution,
our predictions tend to 
overshoot the D0 data (as well as the NLO pQCD predictions) at high $p_T$ values,
but still agree with data within the scale uncertainties.
The latter, since we are working in the LO approximation, are larger compared to the
uncertainties of NLO pQCD calculations, which are of about 8 or 10\%\cite{1,2}.
The evaluation of higher-order corrections in the
$k_T$-factorization approach is the special task and is out of our consideration.
We would like to note, however, that the $k_T$-factorization approach at LO level automatically 
incorporates the main part of the standard (collinear) high-order corrections\cite{11}.
The dependence of our predictions on the unintegrated parton densities is rather
weak: the results coming from the CCFM and KMR parton densities practically coincide
at the Tevatron and are very similar to each other at the LHC conditions.

The relative contributions of gluon-gluon fusion and quark-antiquark annihilation subprocesses 
are shown in Fig.~2, where we used the KMR parton densities for illustration. 
Specially we estimate the contribution of the sea quarks coming from the 
not last gluon splittings (dotted curves) and find it to be of about 10\%. 
Therefore, the uncertainties
connected with the unintegrated parton densities are smaller than the scale uncertainties
of our calculations. To avoid the possible double counting the contribution of the sea quarks coming from the 
earlier gluon evolution steps are not included in the CCFM predictions shown in Fig.~1
since part of them can be already included into the CCFM results (via initial 
parton distributions which enter to the CCFM equation).
Note that these contributions are already included in the quark-antiquark annihilation 
predictions in Fig.~2.

To conclude, in the present note we apply the CCFM and KMR unintegrated parton densities
to the analysis of the first experimental data on the top quark pair
production taken by the CMS and ATLAS collaborations at the LHC
and the recent data taken by the D0 and CDF collaborations at the Tevatron.
Using the off-shell matrix elements of gluon-gluon fusion and quark-antiquark
annihilation subprocesses (where the transverse momentum of both incoming quarks and gluons 
are properly taken into account) we have obtained a reasonably good agreement between our 
predictions and the data. It is important for further studies of small-$x$ physics at hadron colliders, 
and, in particular, for searches of effects of new physics beyond the SM at the LHC.

{\sl Acknowledgements.} 
We are very grateful to 
DESY Directorate for the support in the 
framework of Moscow --- DESY project on Monte-Carlo
implementation for HERA --- LHC. 
A.V.L. was supported in part by the grant of the President of 
Russian Federation (MK-3977.2011.2).
Also this research was supported by the 
FASI of Russian Federation (grant NS-4142.2010.2), 
FASI state contract 02.740.11.0244 and 
RFBR grant 11-02-01454-a.

\newpage

\begin{figure}
\begin{center}
\epsfig{figure=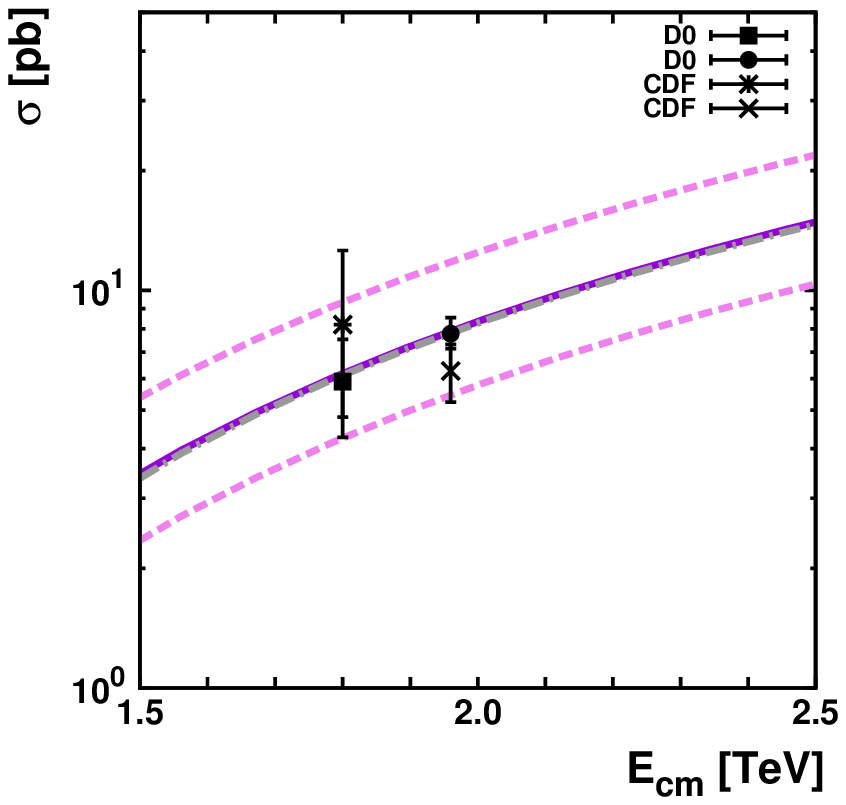, width = 8.1cm}
\epsfig{figure=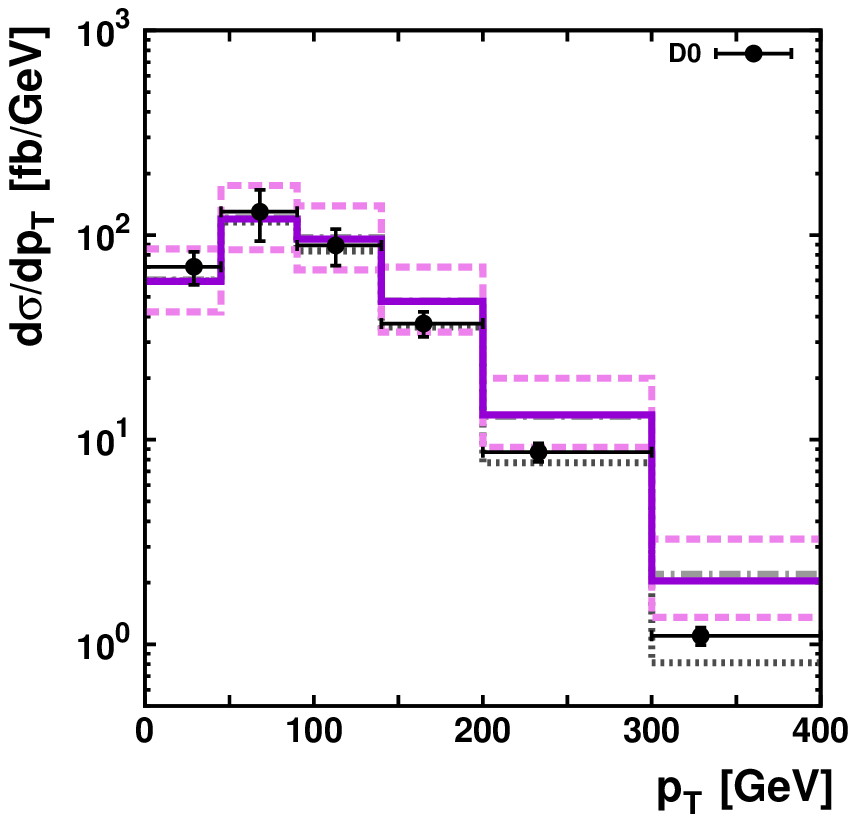, width = 8.1cm}
\epsfig{figure=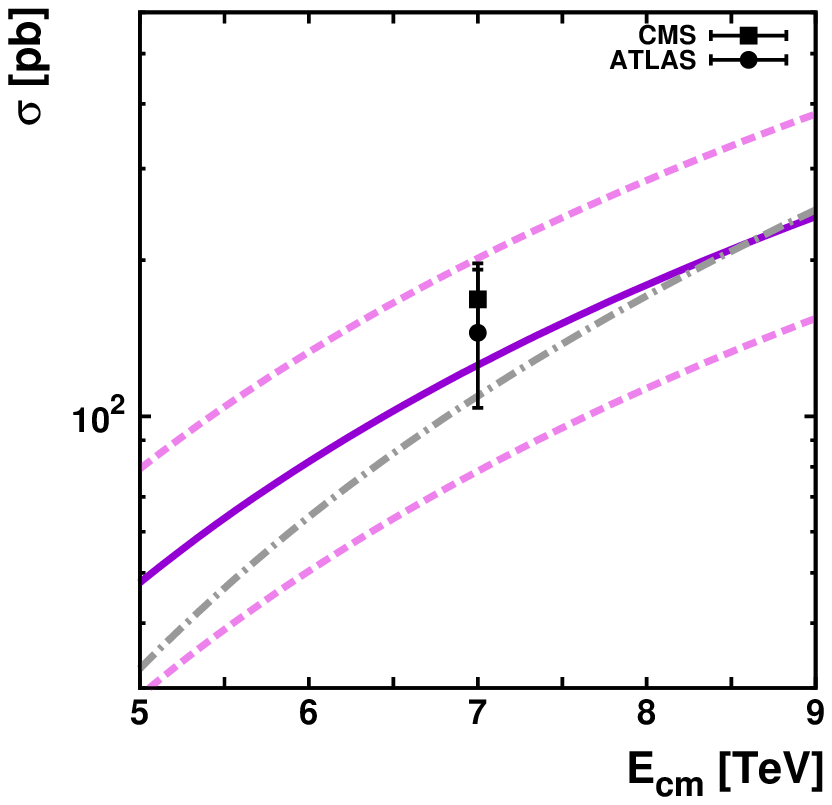, width = 8.1cm}
\caption{The total and differential cross sections of the 
top quark pair production in $p\bar p$ and $pp$ collisions
at the Tevatron and LHC.
The solid and dash-dotted histograms correspond to the results obtained with the KMR and CCFM parton densities, 
respectively.
The upper and lower dashed histograms correspond to scale variations in KMR predictions, as it is
described in the text. The dashed histogram in the transverse momentum distribution 
represents the NLO pQCD predictions\cite{8} listed in\cite{15}.
The experimental data are from CMS\cite{1}, ATLAS\cite{2}, D0\cite{15,16,26} and CDF\cite{17,27}.}
\label{fig1}
\end{center}
\end{figure}

\newpage

\begin{figure}
\begin{center}
\epsfig{figure=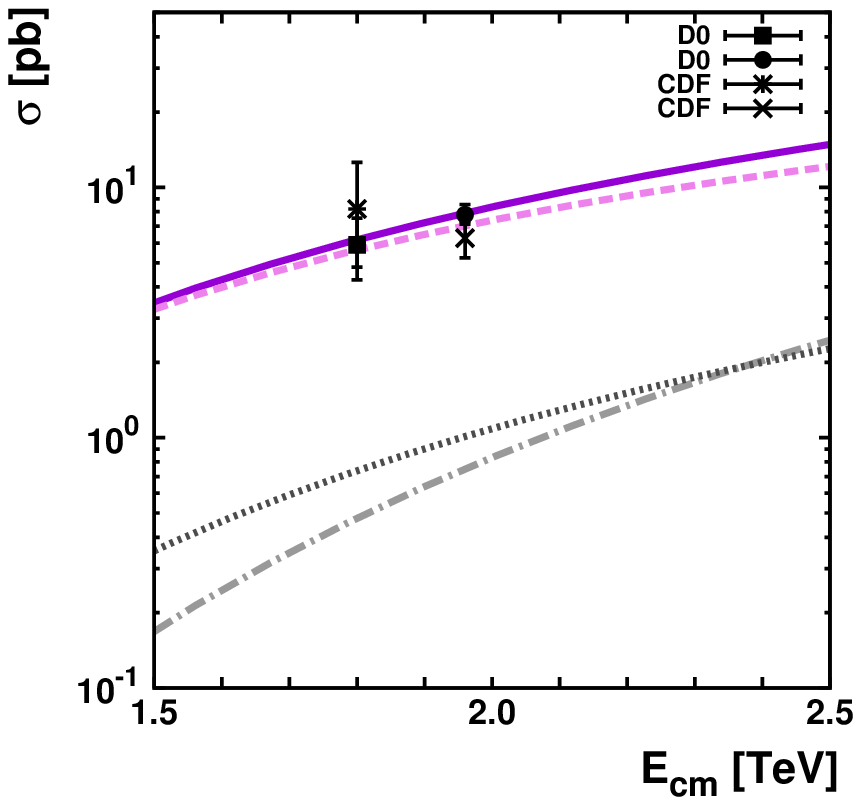, width = 8.1cm}
\epsfig{figure=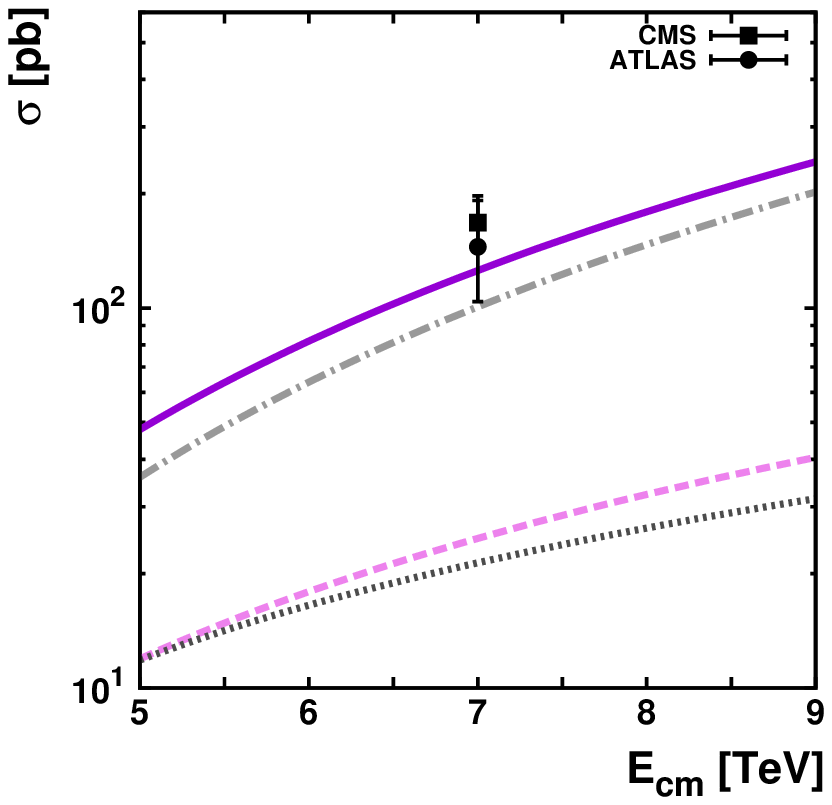, width = 8.1cm}
\caption{Different contributions to the total cross sections of top quark pair 
production in $p \bar p$ and $pp$ collisions at the Tevatron and LHC energies.
The dashed and dash-dotted curves correspond to the
contributions from $q^*q^*\to t\bar t$ and $g^*g^*\to t\bar t$ subprocesses, respectively. 
The solid curve represents the sum of these components.
The dotted curves correspond to the contribution of the sea quarks coming from the 
earlier gluon splittings. 
We use the KMR parton densities for illustration. 
The experimental data are from CMS\cite{1}, ATLAS\cite{2}, D0\cite{15,16,26} and CDF\cite{17,27}.}
\label{fig2}
\end{center}
\end{figure}

\end{document}